# Automatic Identification of Ineffective Online Student Questions in Computing Education

Qiang Hao
Computer Science & SMATE
Western Washington University
Bellingham, WA
qiang.hao@wwu.edu

April Galyardt
Software Engineering Institute
Carnegie Mellon University
Pittsburgh, PA
agalyardt@gmail.com

Bradley Barnes
Computer Science
University of Georgia
Athens, GA
bjb211@uga.edu

Robert Maribe Branch
Learning, Design and Technology
University of Georgia
Athens, GA
rbranch@uga.edu

Ewan Wright
Faculty of Education
University of Hong Kong
Hong Kong, HK
etwright@hku.hk

*Abstract* — This Research Full Paper explores automatic identification of ineffective learning questions in the context of large-scale computer science classes. The immediate and accurate identification of ineffective learning questions opens the door to possible automated facilitation on a large scale, such as alerting learners to revise questions and providing adaptive question revision suggestions. To achieve this, 983 questions were collected from a question & answer platform implemented by an introductory programming course over three semesters in a large research university in the Southeastern United States. Questions were firstly manually classified into three hierarchical categories: 1) learning-irrelevant questions, 2) effective learning-relevant questions, 3) ineffective learning-relevant questions. The inter-rater reliability of the manual classification (Cohen's Kappa) was .88. Four different machine learning algorithms were then used to automatically classify the questions, including Naive Bayes Multinomial, Logistic Regression, Support Vector Machines, and Boosted Decision Tree. Both flat and single path strategies were explored, and the most effective algorithms under both strategies were identified and discussed. This study contributes to the automatic determination of learning question quality in computer science, and provides evidence for the feasibility of automated facilitation of online question & answer in large scale computer science classes.

*Keywords*— online help seeking, automatic question classification, computing education, student questions, learning question classification, educational data mining

## I. Introduction

As the class sizes of computer science (CS) courses grow exponentially each year across the colleges in United States, conventional office hours of instructors can hardly handle all students' learning questions. Therefore, online question & answer (Q & A) platforms have been widely adopted to channel Q & A interactions among students, their peers, and instructors online. Online help seeking through Q & A platforms has special advantages over face-to-face help seeking, such as low threats to self-esteem, easy accessibility, and the choice of anonymity [1, 2].

Despite the advantages, online help seeking poses new challenges to students. Given the asynchronous nature of Q & A interactions, communication is less adaptive, and it usually takes longer to get help. As such, if students fail to ask clear and effective questions, they are not likely to receive helpful answers from either their peers or instructors. The further back-and-forth communication may take even longer and become frustrating. Therefore, it becomes important for students to understand the best way to ask effective questions online.

It is unreasonable to expect instructors to address this challenge by lecturing on help seeking skills, given that they already have a heavy duty of covering all the required learning materials. Besides, relying on individual instructors as the solution is likely to lead to inconsistency, and is difficult to scale up well.

Responding to this challenge, this study explored automatic identification of ineffective learning questions. The immediate and accurate identification of ineffective learning questions opens the door to possible automated facilitation on a large scale, such as alerting learners to revise questions and providing adaptive question revision suggestions. This study contributes to the automatic determination of learning question quality in CS. The results of this study provide evidence for the feasibility of automated facilitation of online Q & A in large scale CS classes.

## II. Literature Review

The traditional thinking on improving students' help seeking skills focused on what factors are most important to help seeking, and investigating proposed intervention strategies on the most significant factors [1, 3, 4]. However, such studies [e.g., 1, 5, 6] were mostly limited to the context of face-to-face help seeking. Besides, such a strategy does not scale up very well to large-scale classes or learners of massive open online course, given the heavy teaching loads of instructors and other constraints. Hao et al. [7] and Hao [8] explored major factors important to face-to-face help seeking on online help seeking, and found that major factors important to face-to-face help seeking, such as achievement goals and interests, were not equally significant to online help seeking, which indicated that either new factors need to be examined or different thinking and solutions are in need to improve students' online help seeking skills.

A series of studies on improving help seeking skills within intelligent tutor systems have been conducted in the last decade [e.g., 9-12]. Despite of their significant findings on how to better interaction between students and tutor systems, help seeking in tutor systems is intrinsically different from in open environments. For most tutor systems examined in prior studies [e.g., 9, 10], helper functions, including on-demands hints and glossaries, were provided side by side of the given learning problems. Students' help seeking in such a closed environment is much simplified. To seek help online in open environments, students need to diagnose problems, organize thinking and languages, and form questions. In contrast, answers or help are just mouse-click away in tutor systems, and all the cognitive labors demanded by open environments are usually not in need when seeking help in tutor systems. Although the studies on help seeking in tutor systems did not directly respond to the challenges of improving students' online help seeking skills, they shed some lights on different thinking of tackling this challenge scalably.

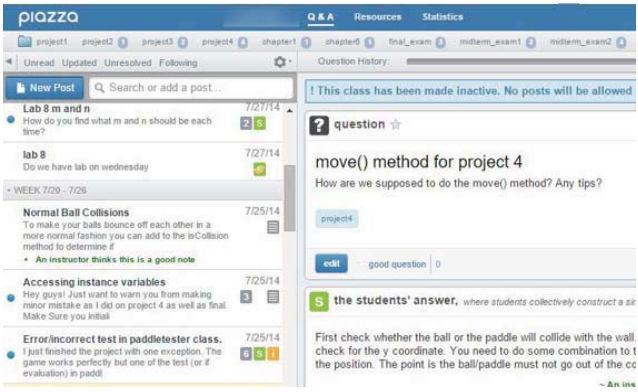

Fig. 1. Interface of Q & A Platform Piazza (https://piazza.com).

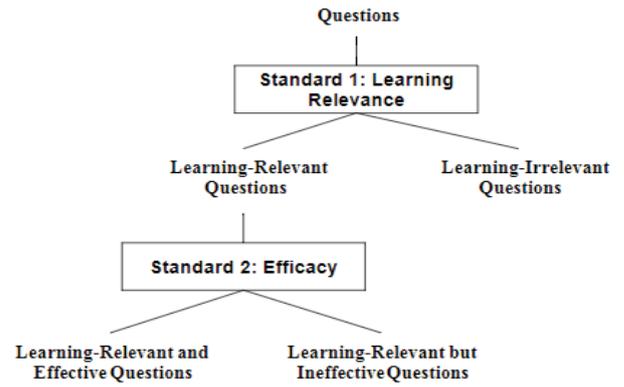

Fig. 2. The Structure of Manual Hierarchical Classification.

Table 1.
CLASSIFICATION RUBRICS OF STUDENTS' QUESTIONS.

| Rubrics | Rationales |
|---|---|
| Whether there is evidence of prior efforts | Prior efforts in problem solving before asking questions are strongly associated with strong capability in self-regulated learning [20]. Questions emerging after problem-solving efforts are more likely to benefit learners and others. Therefore, if there is evidence of prior efforts in a question, it is considered effective. |
| Whether the question is asking for direct answers | Asking for a direct answer is usually deemed as executive help seeking, which indicates a lack of learning desire and wish for expedient task completion [4]. Therefore, a question asking for direct answers is considered ineffective. |
| Whether the question is specific | If a question does not provide sufficient information for others to understand, it is unlikely to solicit helpful answers. Therefore, a question that is not specific enough is considered ineffective. |

Although automatic question classification has been studied extensively, few studies have investigated its application in the field of education. Most studies focused on classifications by topics [e.g., 13-15], semantic functions [e.g., 16, 17] and facts & opinions [e.g., 18, 19]. Few studies have examined automatic question classification from educational perspectives.

## III. RESEARCH PURPOSES

This study explored automatic identification of ineffective learning questions in large-scale CS classes. The two specific questions that guided this study include:

1. How do CS students perform on asking questions online in terms of 1) learning relevance, 2) question efficacy?
2. To what degree of accuracy can we identify CS students' ineffective learning-relevant questions using hierarchical text classification?

## IV. EXPERIMENTS

### A. Data Collection

This study was conducted in an entry-level programming course in a large research university in the southeastern United States. The course typically had around 100 enrolled students every semester. An online Q & A platform, Piazza [see Fig. 1], was implemented in this course for online Q & A interactions. Questions asked on Piazza from students in this course was collected from three semesters across 2015 to 2017. Although the course was taught by different instructors in each semester, the syllabus, course requirements, and course contents remained unchanged during this period.

### B. Manual Classification by Human Experts

In total, 983 questions were collected. The question data come without class information. In order to explore the performance of automatic classification using machine learning algorithms, the collected questions were manually classified firstly. The manual classifications were used as the target values for machine learning algorithms.

The questions were classified hierarchically by trained graduate students majoring in computer science by two standards: 1) learning-relevance, 2) question efficacy (see Fig. 2).

The idea of learning-relevance is self-explanatory. Students may ask questions relevant to learning, but may also ask about due dates of homework or teaching assistants' email addresses. Learning-relevant questions asked in one semester will still be of value to students enrolled in the same course in future semesters. In contrast, learning-irrelevant questions have much less value. As for all learning-relevant questions, we further classified them into effective and ineffective questions. The classification rubrics for question efficacy are presented in Table 1. The coding and classification of a sample question is presented in Table 2.

Table 2.
A SAMPLE OF CODING AND CLASSIFICATION FOR ONE QUESTION.

| Rubrics | |
|---|---|
|  | "I am having a stupid amount of trouble trying [1] to get the code to produce a random word. I'm not really sure what I'm doing wrong [1], but I'm probably not putting the RandomWord.java file into eclipse the right way or something. Can anyone give me some more specific, streamlined instructions [2] for where to download the file and how to call upon the RandomWord.newWord() method? For reference, what I tried to do was simply download the .java file into the src folder of my Goomba java project, and then I tried to retrieve a word with the line: String secretWord = RandomWord.newWord(); Eclipse red underlines RandomWord and suggests that I create class 'RandomWord'. I've tried a few different things but haven't figured anything out, so I decided that I should ask for help here."[3] |
| Has evidence of prior efforts (1) | Yes |
| Is asking for direct answers (2) | No |
| Is specific (3) | Yes |

*This question was classified as effective learning-relevant*

Each question was classified by at least two trained students independently. If a question was classified differently, the classification of that question would be further rated by a third person. The inter-rater reliability (Cohen's Kappa) is .88.

983 questions in total were classified hierarchically into the following three categories:
1. Learning-irrelevant questions
2. Learning-relevant and effective questions
3. Learning-relevant but ineffective questions

743 out of 983 questions were classified as learning-relevant. Among the learning-relevant questions, 366 were classified as effective, and 377 ineffective [see Fig. 3].

*C. Automatic Classification using Machine Learning Algorithms*

Recall of the questions classified as learning-relevant but ineffective was selected as the primary evaluation criteria, and F1-score was selected as the secondary evaluation criteria in this study. The rationale of the evaluation criteria selection are the followings:

1. The primary research goal of this study is to explore whether ineffective learning-relevant questions can be identified, so Q & A platforms can give automatic adaptive suggestions. Therefore, selecting as many ineffective and learning-relevant questions as possible is more of our interest than achieving an overall high accuracy.
2. Imbalance existed in different classes in our results, and F1 score can provide a more comprehensive and combinatory evaluation in such cases.

Four machine learning algorithms were applied to the classifiers, including Naive Bayes Multinomial (NBM), Decision Tree (DT), Logistic Regression (LG), and Support Vector Machines (SVM). Ensemble learning using the

**Manual Question Classification Results**

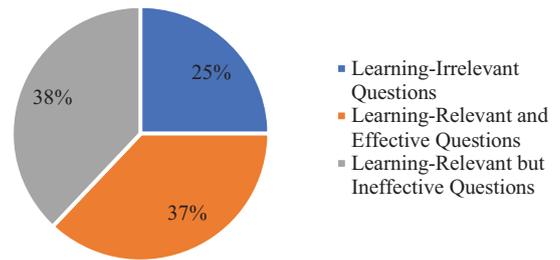

Fig. 3. Manual Question Classification Results.

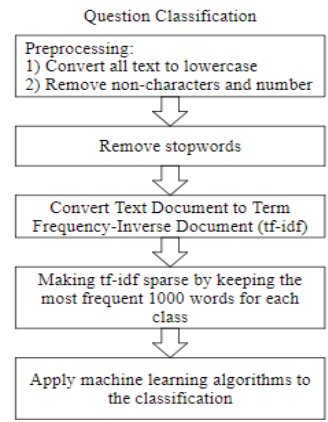

Fig. 4. Process of Automatic Classification.

boosting method was used to stabilize the classification results of Decision Tree, given the inherent instability in its schema. The process of automatic classification using the four machine learning algorithms is presented in Fig. 4. Ten-fold cross validation was used in all classifications to derive the measurement results.

Table 3
MEASUREMENTS OF QUESTION CLASSIFICATION USING FLAT STRATEGY.

| Algorithm | Base-rate Accuracy | Identification of Ineffective Learning-Irrelevant Questions | | |
|---|---|---|---|---|
| | | Recall | Precision | F-Measure |
| SVM | .536 | .864 | .460 | .601 |
| BDT | .532 | .527 | .483 | .504 |
| NBM | .590 | .436 | .526 | .477 |
| LG | .517 | .447 | .459 | .453 |

*Number of all questions: 968, Number of learning-irrelevant questions: 228, Number of effective learning-relevant questions: 364, Number of ineffective learning-relevant questions: 376.*

Table 4
MEASUREMENTS OF CLASSIFICATION ON QUESTION RELEVANCE USING SINGLE PATH STRATEGY.

| Algorithm | Identification of Learning-Relevant Questions | | | |
|---|---|---|---|---|
| | Accuracy | Recall | Precision | F-Measure |
| NBM | .901 | .950 | .923 | .936 |
| SVM | .860 | .932 | .889 | .910 |
| BDT | .840 | .931 | .869 | .899 |
| LG | .789 | .872 | .855 | .863 |

*Number of all questions: 968, Number of learning-irrelevant questions: 228, Number of learning-relevant questions: 740.*

Table 5
MEASUREMENTS OF CLASSIFICATION ON QUESTION EFFICACY USING SINGLE PATH STRATEGY.

| Algorithm | Identification of Learning-Relevant Questions | | | |
|---|---|---|---|---|
| | Accuracy | Recall | Precision | F-Measure |
| SVM | .578 | .847 | .547 | .664 |
| LG | .556 | .547 | .593 | .569 |
| BDT | .531 | .564 | .527 | .545 |
| NBM | .493 | .350 | .563 | .432 |

*Number of all questions: 968, Number of learning-irrelevant questions: 228, Number of learning-relevant questions: 740.*

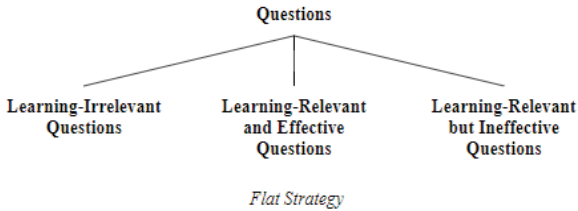
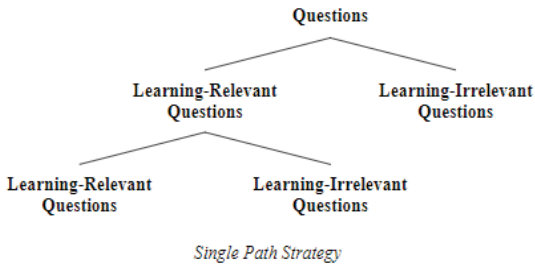

Fig. 5. Application of Flat and Single Path Strategies in Question Classification.

Both flat strategy and single path strategy were used for hierarchical classification in this study. Flat strategy builds a classifier for the all the leaf nodes without considering hierarchical structure [21; see Fig. 5 Top]. The classification results of this strategy are presented in Table 3. SVM is identified as the most effective algorithm by evaluating F1-score. The recall and F1-score of SVM for identifying ineffective learning-relevant questions are .864 and .601 separately.

In contrast, single path strategy builds different classifiers hierarchically at different levels, and allows only one path from the root to a leaf node when testing the classifiers on a question [14; see Fig. 5 Bottom]. Classification algorithms were applied to classification at both relevance and efficacy levels. The results on the first level, learning-relevance, are presented in Table 4. NBM is identified as the most effective algorithm. The recall and F1-score of NBM for identifying learning-relevant questions are .950 and .936 separately.

737 questions were classified as learning-relevant using the model derived from NBM algorithm, including 17 actually irrelevant questions and 720 true relevant questions. The 720 true relevant questions are composed of 360 effective questions and 360 ineffective questions. We further applied the four machine learning algorithms to the 737 questions for efficacy classification. The classification comparisons on question efficacy are presented in Table 5. SVM is identified as the most effective algorithm. The recall and F1-score of SVM are .847 and .664 separately.

V. DISCUSSION

This study made two contributions to automated facilitation of online help seeking in large-scale CS courses. Firstly, we demonstrated the necessity of facilitating online Q & A of entry-level CS students through manual question classification. The ineffective questions were as many as effective questions, which indicated that entry-level students need facilitation in terms asking better learning questions. This result is consistent with literatures that novice CS learners do indeed need guidance and help to improve their help-seeking skills [22-24]. Particularly, there is a pressing need to better understand how to support students in improving their capacity for asking effective questions on online Q & A platforms.

Secondly, the results of this study provide evidence for the feasibility of automated facilitation of online Q & A in large scale CS classes. Both flat and single path strategies were explored in this study to identify ineffective learning questions. The most effective algorithm under the flat strategy was identified as SVM. In comparison, the most effective algorithm combination under the single path strategy was identified as NBM and SVM. In summary, both strategies were able to identify around 85% of all ineffective learning-relevant questions. This is the key step that allows Q & A platforms to provide further adaptive suggestions on question revisions, and the results of this study justified further implementation of adaptive suggestion revision function in Q & A platforms and investigation of its empirical efficacy.

The identification of ineffective questions (Recall: 84.7%; Accuracy Rate: 57.8%) was far from satisfactory. The low accuracy might be due to the definition of effective questions as an umbrella concept composing multiple rubrics. Future studies may consider using the rubrics individually instead of the umbrella efficacy concept to further improve the accuracy.

## VI. Limitations

The present study is not without limitations. First, this study was within one large-scale computer science class. Whether similar results can be found in other large-scale classes and other contexts using the same methods need further investigation. Second, how to further improve the overall accuracy of the automatic classification need to be investigated. Future studies may consider experimenting with a wider range of classification methods, such as ensemble approaches and neural networks. In addition, N-gram features and special stopwords for questions [e.g., 15] may be explored.

## VII. Conclusions

Online help seeking is becoming an increasingly critical skill for computer science students to succeed academically as exponentially increasing number of students show interests in majoring in computer science. This study contributes to the emerging literature on online help seeking of computer science students by (1) proposing a scalable pipeline that facilitates Q & A interaction on Q & A platforms, and (2) demonstrating that results on the automatic question classification using two strategies and four machine learning algorithms. To build on the findings of this study, we call for more research to further investigate the potential of automatic learning question classification.